\documentclass[data,preprints,accept,oneauthor,pdftex]{mdpi} 

\firstpage{1} 
\pubvolume{xx}
\issuenum{1}
\articlenumber{5}
\pubyear{2020}
\copyrightyear{2020}
\history{Received: 6 August 2020; Accepted: 15 September 2020; Published: date}
\updates{yes} 

\usepackage{booktabs}
\usepackage{multirow}
\usepackage{soul}
\usepackage{microtype}

\setitemize{parsep=6pt,itemsep=0pt,leftmargin=*,labelsep=5.5mm}
\setenumerate{parsep=6pt,itemsep=0pt,leftmargin=*,labelsep=5.5mm}
\setlist[description]{itemsep=0mm}

\Title{Bryan's Maximum Entropy Method---Diagnosis of a~Flawed Argument and Its Remedy}

\Author{{Alexander Rothkopf} 
\orcidA{}}

\AuthorNames{Alexander Rothkopf}

\address[1]{%
 Faculty of Science and Technology, University of Stavanger, 4021 Stavanger, Norway; alexander.rothkopf@uis.no}

\corres{Correspondence: alexander.rothkopf@uis.no}

\abstract{The Maximum Entropy Method (MEM) is a popular data analysis technique based on Bayesian inference, which has found various applications in the research literature. While the MEM itself is well-grounded in statistics, I argue that its state-of-the-art implementation, suggested originally by Bryan, artificially restricts its solution space. This restriction leads to a systematic error often unaccounted for in contemporary MEM studies.   The goal of this paper is to carefully revisit Bryan's train of thought, point out its flaw in applying linear algebra arguments to an inherently nonlinear problem, and suggest possible ways to overcome it.}


\begin{document}

\section{Introduction}

Ill-posed inverse problems represent a major challenge to research in many disciplines. They~arise, e.g.,~when one wishes to reconstruct the shape of an astronomical object after its light has passed through a turbulent atmosphere and an imperfect telescope or when we image the interior of a patients skull during a computer tomography scan. In~a more theoretical setting, extracting spectral functions from numerical simulations of strongly correlated quantum fields constitutes another example. The~common difficulty among these tasks lies in the fact that we do not have direct access to the quantity of interest (from here on referred to as $\rho$) but instead only to a distorted representation of it, measured in our experiment (from here on denoted by $D$). Extracting $\rho$ from $D$, in~general, requires us to solve an inherently nonlinear optimization problem, which we construct and discuss in detail~below.
 
Let us consider the common class of inverse problems, where the quantity of interest $\rho$ and the measured data $D$ are related via an integral convolution
\begin{align}
\tilde D(\tau) = \int_{\omega_{\rm min}}^{\omega_{\rm max}} \, d\omega\, K(\tau,\omega) \,\rho(\omega), \; 0\leq\tau\leq\tau_{\rm max}\;,
\end{align}
with a kernel function $K(\tau,\omega)$. For~the sake of simplicity let us assume (as is often the case in practice) that the function $K$ is exactly known. The~task at hand is to estimate the function $\rho$ that underlies the observed $D$. The~ill-posedness (and ill-conditioning) of this task is readily spotted if~we acknowledge that our data comes in the form of $N_\tau$ discrete estimates $D_i=\tilde D(\tau_i)+\eta$ of the true function $\tilde D$, where $\eta$ denotes a source of noise. In~addition, we need to approximate the integral in some form for numerical treatment. In~its simplest form, writing it as a sum over $N_\omega$ bins we obtain
\begin{align}
D_i = \sum_{l=1}^{N_\omega} \Delta\omega_l K_{il} \,\rho_l. \label{Eq:convdisc}
\end{align}
At this point, we are asked to estimate $N_\omega \gg N_\tau$ optimal parameters $\rho_l$ from $N_\tau$ data points, which themselves carry uncertainty.  A~naive $\chi^2$ fit of the $\rho_l$'s is of no use, since it would produce an infinite number of degenerate solutions, which all reproduce the set of $D_i$'s within their error bars. Only~if we introduce an appropriate regularization can the problem be made well-posed, and it is this regularization which in general introduces~nonlinearities.

Bayesian inference represents one way to regularize the inversion task. It provides a systematic procedure for how additional (so called prior) knowledge can be incorporated to that effect. Bayes theorem
\begin{align}
P[\rho|D,I] = \frac{P[D|\rho,I]P[\rho|I]}{P[D|I]},
\end{align}
states that the {\it posterior} probability $P[\rho|D,I] $ for some set of parameters $\rho_l$ to~be the true solution of the inversion problem is given by the product of the {\it likelihood} probability $P[D|\rho,I]$ and the {\it prior} probability $P[\rho|I]$. The~$\rho$ independent normalization $P[D|I]$ is often referred to as the {\it evidence}. Assuming that the noise $\eta$ is Gaussian, we may write the likelihood as
\begin{align}
&P[D|\rho,I]={\rm exp}[-{\cal L}], \qquad {\cal L}=\frac{1}{2} \sum_{ij} \big( D_i - D^{\rho}_i \big) C_{ij}^{-1} \big( D_j- D^{\rho}_j \big), \label{Eq:Likelihood}
\end{align}
where $C_{ij}$ is the unbiased covariance matrix of the measured data with respect to the true mean and $D^\rho_i$ refers to the synthetic data that one obtains by inserting the current set of $\rho_l$ parameters into \eqref{Eq:convdisc}. A~$\chi^2$ fit would simply return one of the many degenerate extrema of ${\cal L}$, hence being referred to as a maximum likelihood~fit.

The important ingredient of Bayes theorem is the presence of the prior probability, often expressed in terms of a regulator functional ${\cal S}$
\begin{align}
&P[\rho|I]=\rm exp[{\cal S}].
\end{align}

It is here where pertinent domain knowledge can be encoded. For~the study of intensity profiles of astronomical objects and hadronic spectral functions for example, it is~a-priori known that the values of $\rho$ must be positive. Depending on which type of information one wishes to incorporate, the~explicit form of ${\cal S}$ will be different. It is customary to parameterize the shape of the prior distribution by two types of hyperparameters, the~default model $m(\omega)$ and a confidence function $\alpha(\omega)$. The~discretized $m_l=m(\omega_l)$ represents the maximum of the prior $P[\rho_l | I(m,\alpha)]$ for each parameter $\rho_l$  and each $\alpha_l=\alpha(\omega_l)$ its corresponding~spread.

Once both the likelihood and prior probability are set, we may search for the maximum a posteriori (MAP) point estimate of the $\rho_l$'s via
\begin{align}
\left.\frac{\delta }{\delta \rho} P[\rho|D,I]\right|_{\rho=\rho^{\rm Bayes}}=0 \label{Eq:MAPest}.
\end{align}
$\rho^{\rm Bayes}$ constitutes the most probable parameters given our data and prior knowledge. (In the case that no data is provided, the Bayesian reconstruction will simply reproduce the default model $m$.) Note that since the exponential function is monotonous, instead of finding the extremum of $P[\rho|D,I]$, in~practice, one often considers the extremum of ${\cal L}-{\cal S}$ directly.

The idea of introducing a regularization in order to meaningfully estimate the most probable set of parameters underlying observed data has a long history. As~early as 1919~\cite{whittaker_1922}, it was proposed to combine what we here call the likelihood with a smoothing regulator. Let us have a look at three choices of regulators from the literature: the historic Tikhonov (TK) regulator~\cite{TIKHONOV:1943} (1943--), the~Shannon--Jaynes entropy deployed in the Maximum Entropy Method (MEM)~\cite{MEM:1986,Jarrell:1996rrw} (1986--), and the more recent Bayesian Reconstruction (BR) method~\cite{Burnier:2013nla} regulator (2013--), respectively,
\begin{align}
&{\cal S}_{\rm TK} = -\sum_l \Delta\omega_l \, \alpha_l \,\frac{1}{2}\big( \rho_l - m_l \big)^2, \label{eq:TK}\\
&{\cal S}_{\rm MEM} = \sum_l \Delta\omega_l \, \alpha_l \,\big( \rho_l - m_l - \rho_l {\rm log}\big[ \frac{\rho_l}{m_l}\big] \big), \label{eq:SJ} \\
&{\cal S}_{\rm BR} = \sum_l \Delta\omega_l \, \alpha_l \,\big( 1-\frac{\rho_l}{m_l} + {\rm log}\big[ \frac{\rho_l}{m_l}\big]  \big). \label{eq:BR}
\end{align}

Both ${\cal S}_{\rm MEM} $ and ${\cal S}_{\rm BR} $ are axiomatically constructed, incorporating the assumption of positivity of the function $\rho$. The~assumption manifests itself via the presence of a logarithmic term that forces $\rho$ to be positive-semidefinite in the former and positive-definite in the latter case. It is this logarithm that is responsible for the numerical optimization problem \eqref{Eq:MAPest} to become genuinely~nonlinear.

Note that all three functions are concave, which (as proven for example~in~\cite{Asakawa:2000tr}) guarantees that if an extremum of $P[\rho|D,I]$ exists, it is unique---i.e., within the $N_\omega$ dimensional solution space spanned by the discretized parameters $\rho_l$, in~the case that a Bayesian solution exists, we will be able to locate it with standard numerical methods in a straightforward~fashion.

\section{Diagnosis of the Problem in Bryan's~MEM}

In this section, we investigate the consequences of the choice of regularization on the determination of the most probable spectrum. Starting point is the fully linear Tikhonov regularization, on~which a lot of the intuition in the treatment of inverse problems is built. We then continue to the Maximum Entropy Method, which amounts to a genuinely nonlinear regularization and point out how arguments, which~were valid in the linear case, fail in the nonlinear~context.

\subsection{Tikhonov~Regularization}

The Tikhonov choice amounts to a Gaussian prior probability, which allows $\rho$ to take on both positive and negative values. The~default model $m_l$ denotes the value for $\rho_l$, which was most probable before the arrival of the measured data $D$ (e.g., from a previous experiment), and $\alpha_l$ represents our confidence into the prior knowledge (e.g., the uncertainty of the previous experiment). 

Since both \eqref{Eq:Likelihood} and \eqref{eq:TK} are at most quadratic in $\rho_l$, taking the derivative in \eqref{Eq:MAPest} leads to a set of linear equations that need to be solved to compute the Bayesian optimal solution $\rho^{\rm Bayes}$. It is this fully linear scenario from~which most intuition is derived when it comes to the solution space of the inversion task. Indeed, we are led to the following relations
\begin{align}
&-\alpha_l (\rho_l - m_l) = \sum_{i} K_{il}\frac{\delta {\cal L}}{\delta D^\rho_i}, \quad  -\hat \alpha ( \vec{\rho}- \vec{m} ) = \hat K^T \vec{ \frac{\delta {\cal L}}{\delta D^\rho}},\label{eq:TKoptcrit}
\end{align}
which can be written solely in terms of linear vector-matrix operations. Note that in this case $\delta {\cal L}/{\delta D^\rho}$ contains the vector $\vec{\rho}$ in a linear fashion. \eqref{eq:TKoptcrit} invites us to parameterize the function $\rho$ via its deviation from the default model
\begin{align}
\vec{\rho}=\vec{m}+\vec{a},
\end{align}
and to look for the optimal set of parameters $a_l$. Here, we may safely follow Bryan~\cite{bryan_maximum_1990} and investigate the singular values of $K^T = U \Sigma V^t$ with $U$ being an $N_\omega\times N_\omega$ special orthogonal matrix, $\Sigma$ an $N_\omega\times N_\tau$ matrix with $N_\tau$ nonvanishing  diagonal entries, corresponding to the singular values of $K^T$ and  $V^t$ being an $N_\tau\times N_\tau$ special orthogonal matrix.  
We are led to the expression
\begin{align}
& -\hat \alpha  \vec{a}  = \hat U \hat \Sigma {\hat V}^t \vec{ \frac{\delta {\cal L}}{\delta D^\rho}},
\end{align}
which tells us that in this case, the solution of the Tikhonov inversion lies in a functional space spanned by the first $N_\tau$ columns of the matrix $\hat U$ (usually referred to as the SVD or singular subspace) around the default model $\vec{m}$---i.e., we can parameterize $\rho$ as
\begin{align}
\vec{\rho}=\vec{m}+\sum_{k=1}^{N_\tau} c_k \vec{U}_k. \label{eq:TKparm}
\end{align}

The point here is that if we add to this SVD space parametrization any further column of the matrix $\hat U$, it directly projects into the null space of $K$ via $\hat K \cdot \vec{U}_{k>N_\tau} =0$. In~turn, such a column does not  contribute to computing synthetic data via \eqref{Eq:convdisc}. As~was pointed out in~\cite{Asakawa:2020hjs}, in such a linear scenario, the~SVD subspace is indeed all there is. If~we add extra columns of $\hat U$ to the parametrization of $\rho$, these do not change the likelihood. Thus, the corresponding parameter $c_j$ of that column is not constrained by data and will come out as zero in the optimization procedure of \eqref{Eq:MAPest}, as~it encodes a deviation from the default model, which is minimized by ${\cal S}$.

\subsection{Maximum Entropy~Method}

Now that we have established that in a fully linear context the arguments based on the SVD subspace are indeed justified, let us continue to the Maximum Entropy Method, which deploys the Shannon--Jaynes entropy as regulator. ${\cal S}_{\rm MEM}$ encodes as prior information, e.g.,~the positivity of the function $\rho$, which manifests itself in the presence of the logarithm in \eqref{eq:SJ}. This logarithm however also entails that we are now dealing with an inherently nonlinear optimization problem in \eqref{Eq:MAPest}. Carrying out the functional derivative with respect to $\rho$ on ${\cal L}-{\cal S}$, we are led to the following relation
\begin{align}
-\alpha_l {\rm log}\big[\frac{\rho_l}{m_l}\big] = \sum_{i} K_{il}\frac{\delta {\cal L}}{\delta D^\rho_i}. \label{eq:Bryan1}
\end{align}

As suggested by Bryan~\cite{bryan_maximum_1990}, let us introduce the completely general (since $\rho>0$) and nonlinear redefinition of the parameters
\begin{align}
\rho_l=m_l{\rm exp}[a_l]  \label{eq:Bryan2}.
\end{align}

Inserting \eqref{eq:Bryan2} into \eqref{eq:Bryan1}, we are led to an expression that is formally quite similar to the result obtained in the Tikhonov case
\begin{align}
&-\alpha_l a_l = \sum_{i} K_{il}\frac{\delta {\cal L}}{\delta D^\rho_i}, \qquad -\hat \alpha \vec{a} = \hat K^T \vec{ \frac{\delta {\cal L}}{\delta D^\rho}}.\label{eq:SJoptcrit}
\end{align}

While at first sight this relation is also amenable to be written as a linear relation for $\vec{a}$, it is actually fundamentally different from \eqref{eq:TKoptcrit}, since due to its nonlinear nature, $\vec{a}$ enters $\delta {\cal L}/{\delta D^\rho}$ via componentwise exponentiation. It is here, when attempting to make a statement about such a nonlinear relation with the tools of linear algebra, that we run into difficulties. What do I mean by that? Let us push ahead and introduce the SVD decomposition of the transpose Kernel as before,
\begin{align}
-\hat \alpha \vec{a} = \hat U \hat \Sigma {\hat V}^t \vec{ \frac{\delta {\cal L}}{\delta D^\rho}}.
\end{align}

At first sight, this relation seems to imply that the vector $\vec{a}$---which encodes the deviation from the default model (this time multiplicatively)---is restricted to the SVD subspace, spanned by the first $N_\tau$ entries of the matrix $\hat U$. My claim (as put forward most recently in~\cite{Rothkopf:2019ipj}) is that this conclusion is false, since this linear-algebra argument is not applicable when working with \eqref{eq:SJoptcrit}.  Let us continue to setup the corresponding SVD parametrization advocated, e.g.,~in~\cite{Asakawa:2020hjs}
\begin{align}
\rho_l=m_l{\rm exp}\big[ \sum_{k=1}^{N_\tau} c_k U_{lk} \big]. \label{eq:SJparm}
\end{align}

In contrast to \eqref{eq:TKparm}, the SVD space is not all there is to \eqref{eq:SJparm} (see explicit computations in Appendix~\ref{sec:SVDex}). This we can see by taking the $N_{\tau}+1$ column of the matrix $\hat U$, exponentiating it componentwise, and applying it to the matrix $K$. In~general, we get
\begin{align}
\sum_{l}\; K_{il} {\rm exp}\big[U_{l(N_{\tau}+1)}\big] \neq 0. \label{eq:nonnull}
\end{align}

This means that if we add additional columns of the matrix $\hat U$ to the parametrization in \eqref{eq:SJparm}, they do not automatically project into the null-space of the Kernel (see the explicit example in Appendix~\ref{sec:SVDex}
of this manuscript) and thus will contribute  to the likelihood. In~turn, the corresponding parameter $c_j$ related to that column will not automatically come out to be zero in the minimization procedure~\eqref{Eq:MAPest}. Hence, we cannot a priori disregard its contribution and thus, the contribution of this direction of the search space, which is not part of the SVD subspace. We thus conclude that limiting the solution space in the MEM to the singular subspace amounts to an ad-hoc procedure, motivated by an incorrect application of linear-algebra arguments to a fully nonlinear optimization~problem.

A representative example from the literature, where the nonlinear character of the parametrization of $\rho$ is not taken into account, is the recent~\cite{Asakawa:2020hjs} (see Equations~(7) and (12) in that manuscript). We emphasize that one does not apply a column of the matrix $\hat U$ itself to the matrix K but the componentwise exponentiation of this column. This operation does not project into the~null-space.

In those cases where we have only few pieces of reliable prior information and our datasets are limited, restricting to the SVD subspace may lead to significantly distorted results (as shown explicitly in~\cite{Rothkopf:2011ef}). On~the other hand, its effect may be mild if the default model already encodes most of the relevant features of the final result and the number of datapoints is large, so that the SVD subspace is large enough to (accidentally) encompass the true Bayesian solution sought after in \eqref{Eq:MAPest}. Independent of the severity of the problem, the~artificial restriction to the SVD subspace in Bryan's MEM is a source of systematic error, which needs to be accounted for when the MEM is deployed as precision tool for inverse~problems.

Being liberated from the SVD subspace does not lead to any conceptual problems either. We~have brought to the table $N_\tau$ points of data, as~well as $N_\omega$ points of prior information in the form of the default model $m$ (as well as its uncertainty $\alpha$). This is enough information to determine the $N_\omega$ parameters $\rho_l$ uniquely, as~proven in~\cite{Asakawa:2000tr}. 

Recognizing that linear-algebra arguments fail in the MEM setup also helps us to understand some of the otherwise perplexing results found in the literature. If~the singular subspace were all there is to the parametrization of $\rho_l$ in \eqref{eq:SJparm}, then it would not matter whether we use the first $N_\tau$ columns of $\hat U$ or just use the $N_\tau$ columns of $\hat K^T$ directly. Both encode the same target space, the~difference being only that the columns of $\hat U$ are orthonormal. However, as~was clearly seen in Figure~28 of~\cite{Jakovac:2006sf}, using the SVD parametrization or the columns of $K^T$ leads to significantly different results in the reconstructed features of $\rho$. If~the MEM were a truly linear problem, these two parameterizations gave exactly the same result. The~finding that the results do not agree emphasizes that the MEM inversion is genuinely nonlinear and the restriction to the SVD subspace is~ad hoc.

\subsection{Numerical Evidence for the Inadequacy of the SVD~Subspace}

Let us construct an explicit example to illustrate the fact that the solution of the MEM reconstruction may lie outside of the SVD search space. Since Bryan's derivation of the SVD subspace proceeds independently of the particular form of the kernel $K$, the~provided data $D$, and the choice of the default model $m$, we are free to choose them at will. For~our example, we consider a transform often encountered among inverse problems related to the physics of strongly correlated quantum~systems.
 
One then has $K(\tau,\omega)=1/(\omega^2 +\tau^2)$ and we may set $m=1$. With~$\alpha$ entering simply as scaling factor in \eqref{eq:SJoptcrit}, we do not consider it further in the following. Let me emphasize again that the arguments leading to \eqref{eq:SJoptcrit} did not make any reference to the data we wish to reconstruct. Here, we will consider three datapoints that encode a single delta peak at $\omega=\omega_0$, embedded in a flat~background.

\begin{figure}[H]
\centering 
\includegraphics[scale=0.42]{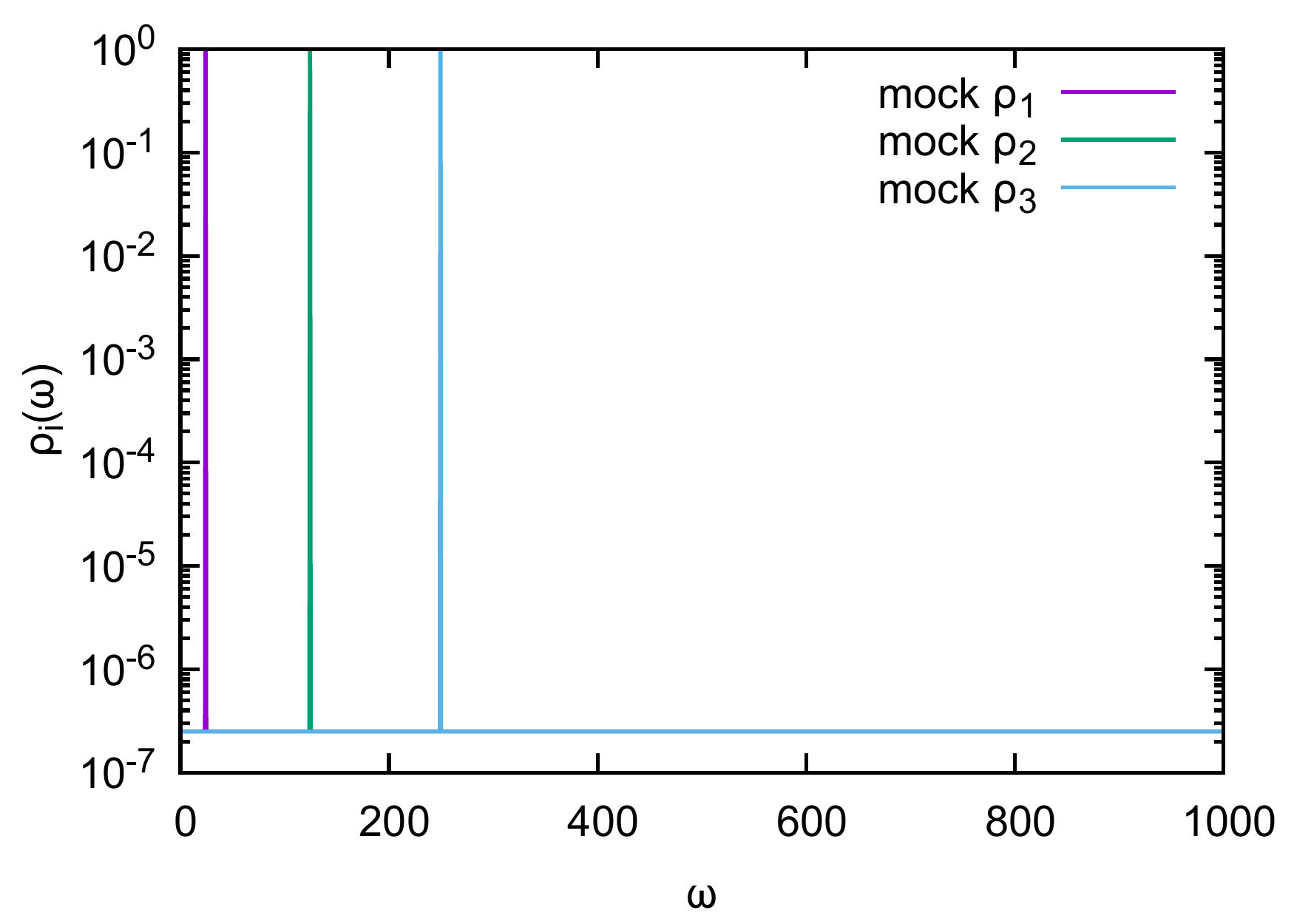}
\includegraphics[scale=0.42]{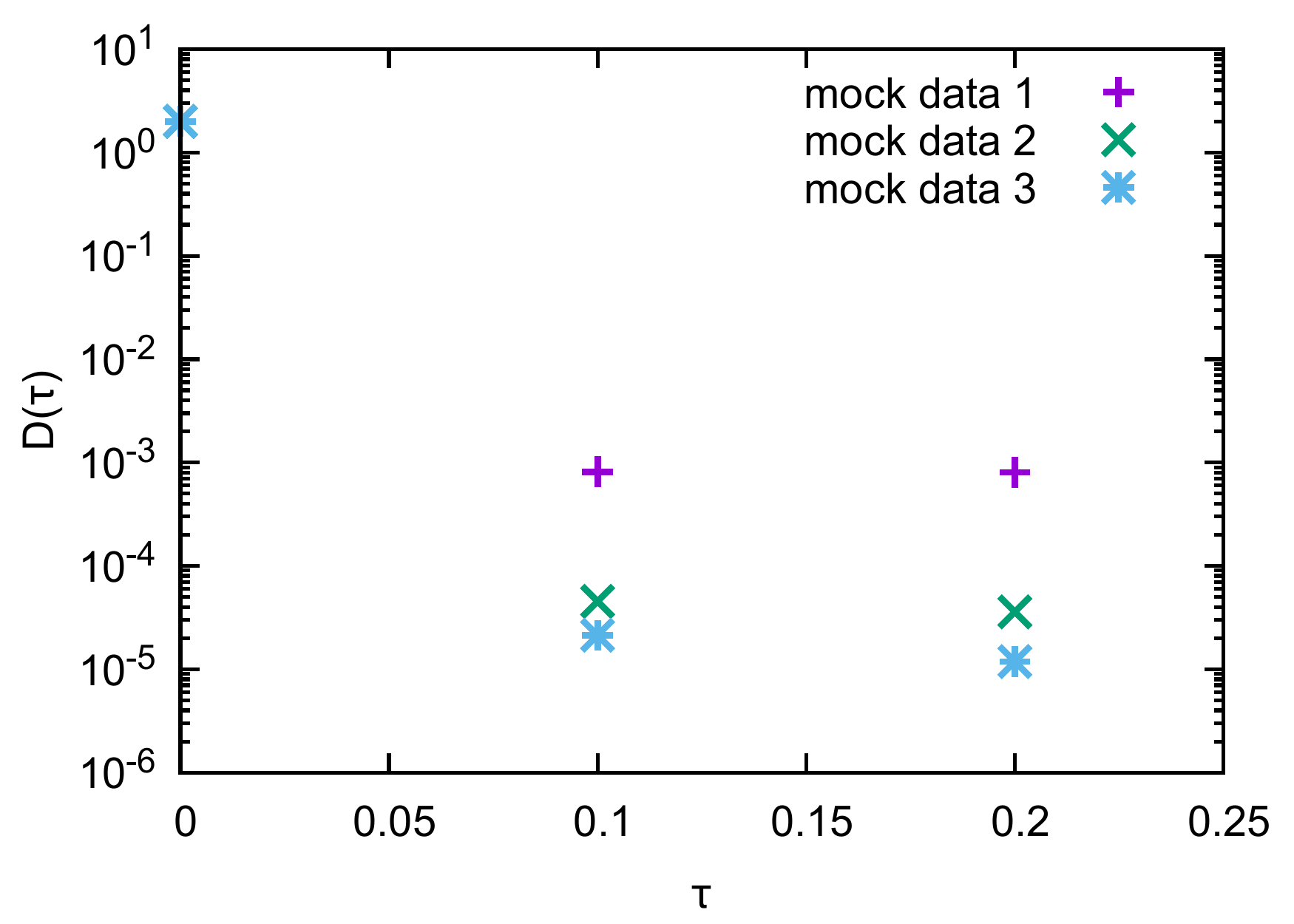}
\caption{{The mock function} 
$\rho$ (left) and corresponding mock data (right) deployed in the explicit example discussed in the main text, which shows that the extremum of the posterior is not necessarily located within the singular value decomposition (SVD)~subspace. }\label{Fig:mockspecdat}
\end{figure}

Now, let us discretize the frequency domain between $\omega_{\rm min}=1/2000$ and $\omega_{\rm max}=1000$ with $N_\omega=2000$ points. Together with the choice of $\tau_{\rm min}=0$, $\tau_{\rm max}=0.2$, and $N_\tau=3$, this fully determines the kernel matrix $K_{il}$ in \eqref{Eq:convdisc}. Three different mock functions $\rho_i$ are considered, with~$\omega_0^{(1)}=25$, $\omega_0^{(2)}=125$, and $\omega_0^{(3)}=250$, the~background is assigned the magnitude $1/N_\omega^2$ (see Figure~\ref{Fig:mockspecdat} left).

\begin{figure}[H]
\centering
\includegraphics[scale=0.42]{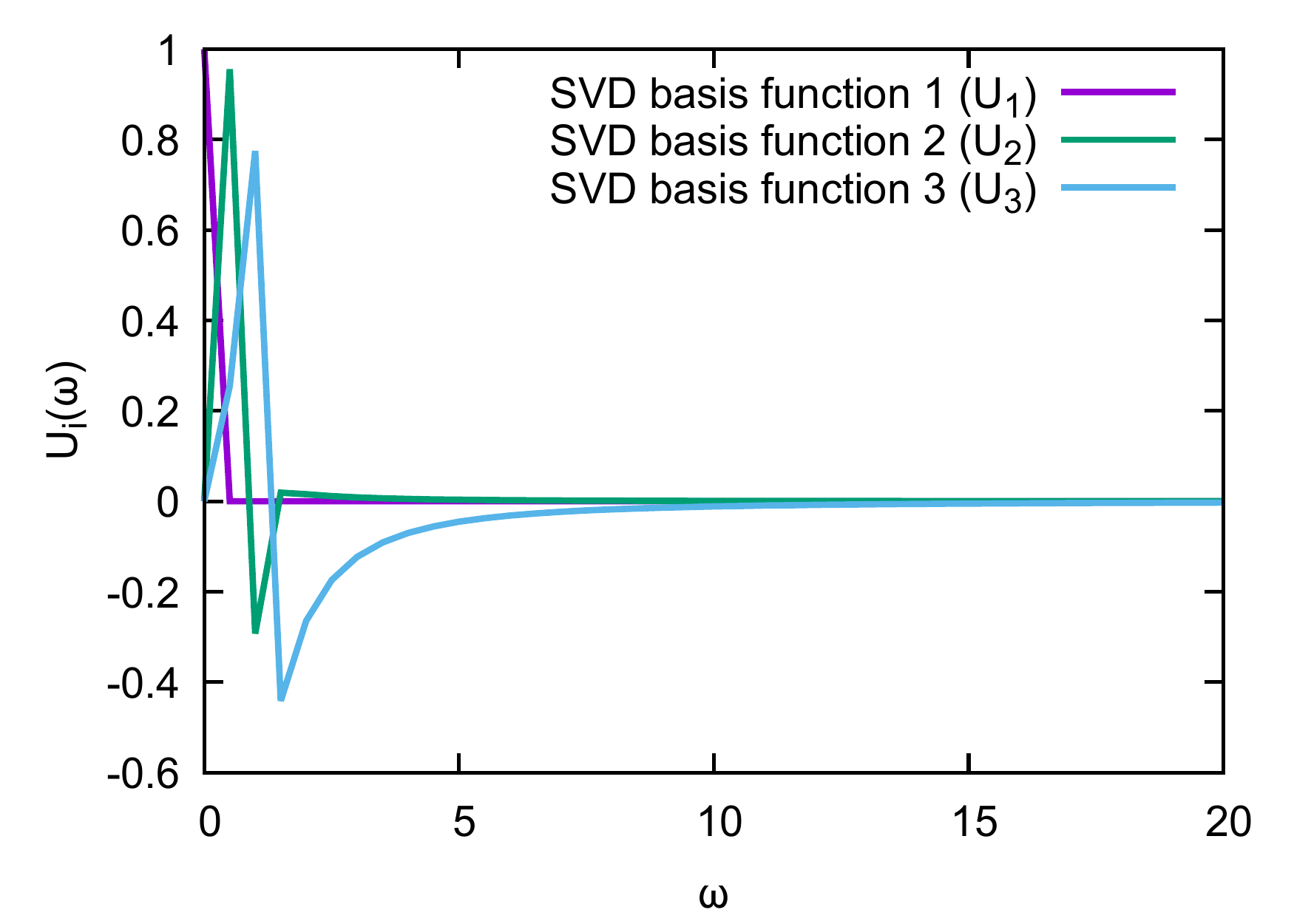}
\caption{{The SVD basis functions} $U_1(\omega)$, $U_2(\omega)$, and $U_3(\omega)$, which Bryan's argument suggests should capture the extremum of the posterior for all three mock functions $\rho_i$ after exponentiation. Note that the SVD basis functions flatten out about $\omega=10$. Their domain extends to $\omega_{\rm max}=1000$ but their value stays close to~zero.}\label{Fig:mockspecdat2}
\end{figure}

Bryan's argument states that in the presence of three datapoints (see Figure~\ref{Fig:mockspecdat} right), irrespective of the data encoded in those datapoints, the~extremum of the posterior must lie in the space spanned by the exponentiation of the three first columns of the matrix $\hat U$, obtained from the SVD of the transpose kernel $K^t$. In~Figure~\ref{Fig:mockspecdat2}, its first three columns are explicitly plotted. Note that while they do show some peaked behavior around the origin, they quickly flatten off above $\omega=10$. From~this inspection by eye, it already follows that it will be very difficult to linearly combine  $U_1(\omega)$, $U_2(\omega)$, and $U_3(\omega)$ into a sharply peaked function, especially for a peak located at $\omega>10$.

Assuming that the data comes with constant relative error $\Delta D/D=\kappa$, let us find out how well we can reproduce it within Bryan's search space. A~minimization carried out by Mathematica (see~the explicit code in Appendix \ref{sec:MEMex}) tells us that ${\cal L}_{\rm min}^1\approx 10^6 \kappa^{-2}$, ${\cal L}_{\rm min}^2\approx 10^9 \kappa^{-2}$, and ${\cal L}_{\rm min}^3 \approx 10^{10}\kappa^{-2}$---i.e.,~we~clearly see that we are not able to reproduce the provided datapoints well (i.e., $\chi^2/{\rm d.o.f}={\cal L}/3\gg1$) and that as the deviation becomes more and more pronounced, the~higher the delta peak is positioned along $\omega$. In~the full search space on the other hand, we can always find a set of $\rho_l$'s which bring the likelihood as close to zero as~desired.

Minimizing the likelihood of course is not the whole story in a Bayesian analysis, which is why we have to take a look at the contributions from the regulator term ${\cal S}$. Remember that by definition, it is negative. We find that for all three mock functions $\rho_i$, Bryan's best fit of the likelihood leads to values of ${\cal S}\approx -0.24$.  On~the other hand, a value of ${\cal S}\approx -200$ is obtained in the full search space for the parameter set $\rho_l$ which contains only one entry that is unity and all other entries being set to the background at $1/N_\omega^2$.

From the above inspection of the extremum of ${\cal L}$ and the associated value of ${\cal S}$ inside and outside of Bryan's search space, we arrive at the following: Due to the very limited structure present in the SVD basis functions $U_1(\omega)$, $U_2(\omega)$, and $U_3(\omega)$, it is in general not possible to obtain a good reconstruction of the input data. This in turn leads to a large minimal value of ${\cal L}$ accessible within the SVD subspace. Due to the fact that ${\cal S}$ cannot compensate for a large value of ${\cal L}$, we have constructed an explicit example where at least one set of $\rho_l$'s (the one that brings ${\cal L}$ close to zero in the full search space) leads to a smaller value of ${\cal L}-{\cal S}$ and thus to a larger posterior probability than any of the $\rho_l$'s within the SVD~subspace.

In other words, we have constructed a concrete example in which the MEM solution, given by the global extremum of the posterior, is not contained in the SVD~subspace.

\section{Remedy of the~Problem}

Having identified the shortcoming of Bryan's MEM as the ad-hoc restriction of the search space to the SVD subspace, we must consider how to overcome it. In~this section, we present two possible routes to do so. In~the first subsection, systematic modifications of the MEM, previously put forward by the author, are considered, which extend and generalize the SVD search space. In~the second subsection, we present a collection of modern approaches to Bayesian inference developed by the research community which are not affected by limited search spaces. Particular attention is paid to the BR method recently codeveloped by the~author.

\subsection{Staying within~MEM}

The most naive path to take is to simply add additional columns of the matrix $\hat U$ to the nonlinear parametrization of \eqref{eq:SJparm}, as~proposed by the author in 2011 in~\cite{Rothkopf:2011ef} and successfully applied in practice in~\cite{Rothkopf:2011db}. This procedure systematically approaches the full search space, in~which the unique Bayesian solution is located ({{We are surprised by the statements} made recently in~\cite{Asakawa:2020hjs}, which claim that the numerics presented in~\cite{Rothkopf:2011ef} and thus also in~\cite{Rothkopf:2011db} were unreliable. Both papers have been peer-reviewed and the underlying code has been freely available for many years on the author's website (\href{http://www.alexrothkopf.de/files/ExtMEMv3.tar.bz2}{ExtMEM link}). Furthermore, the examples presented in~\cite{Rothkopf:2011ef} are explicitly parameterized and thus lend themselves to straightforward reproduction.}). 

Investigating in detail the shape of the SVD basis functions, it was found that they may not offer the same resolution for features located at different $\omega$'s. In~practice, this means that one often needs to add a large number of additional SVD basis functions before the structures encoded in the data $D_i$ are sufficiently resolved. In~order to remedy this situation, we can exploit that we are not bound by the SVD subspace and instead, as~suggested in~\cite{Rothkopf:2012vv}, deploy for example~Fourier basis functions in \eqref{eq:SJparm}---i.e.,~we replace the columns of $\hat U$ by a linear combination of $sin$ and $cos$ terms. This proposal has been shown to lead to a resolution capacity of the MEM that is independent of the position of the encoded structures in $\rho$ along $\omega$ and has been put in practice, e.g.,~in~\cite{Kelly:2018hsi}.

Thanks to the availability of highly efficient numerical optimization algorithms, such as the limited-memory BFGS (Broyden–Fletcher–Goldfarb–Shanno) algorithm, it is nowadays easily possible to directly carry out the optimization task \eqref{Eq:MAPest} in the full $N_\omega$ dimensional search space, even if $N_\omega\sim {\cal O}(10^3)$. Together with the proof of uniqueness of the Bayesian solution and the related convexity of ${\cal L}-{\cal S}$, there does not exist any principal need to restrict to a low-dimensional~subspace.

\subsection{Beyond~MEM}

An active community is working on devising improved Bayesian approaches to inverse problems in the sciences. On~the one hand, there exist works that go beyond the maximum a-posteriori approach and proceed towards sampling the posterior, such as the stochastic analytic continuation method~\cite{Sandvik:1998}, of~which the MEM is but one special limit as shown in~\cite{Beach:2004}. Together with the SOM method presented in~\cite{mishchenko2000}, these stochastic methods have for example,~been deployed in the study of nuclear matter at high temperatures in~\cite{Ding:2017std}. Recently, the community has seen heightened activity in exploring the use of neural networks for the solution of inverse problems, e.g.,~in~\cite{yoon_analytic_2018,fournier_artificial_2018,Kades:2019wtd,Offler:2019eij}.

\begin{figure}[t]
\centering
\includegraphics[scale=0.6]{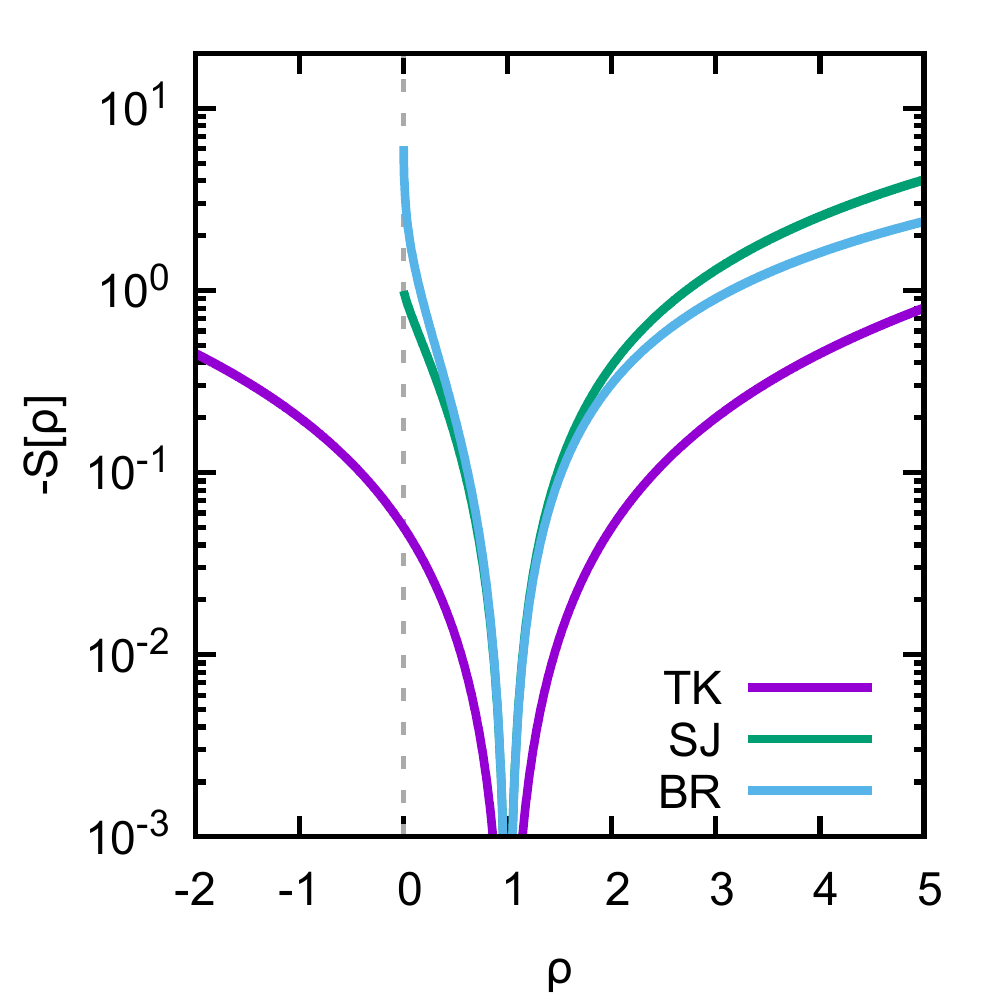}
\includegraphics[scale=0.6]{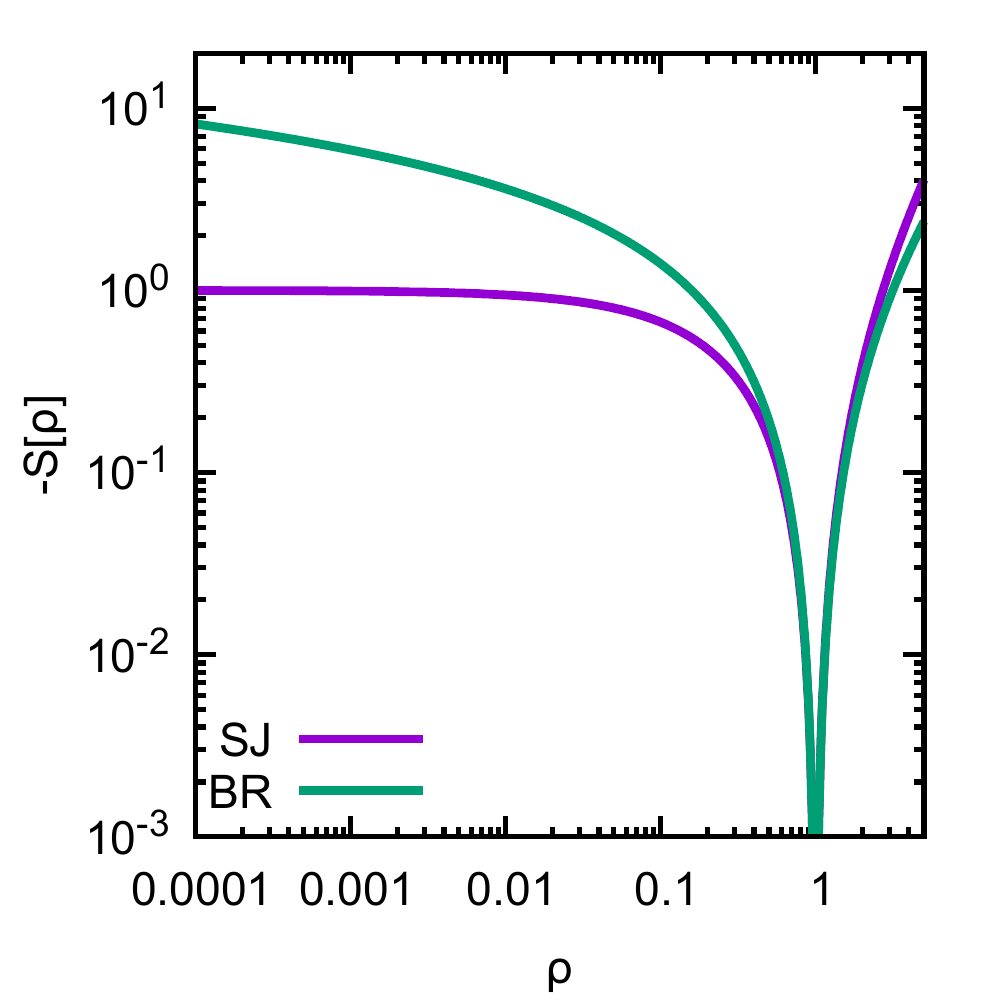}
\caption{{Comparison of the behavior} of the integrand of the Tikhonov (TK), MEM, and Bayesian Reconstruction (BR) regulators for the choice $\alpha=0.1$ and $m=1$ using (left) a log-lin scale and (right) a  log-log scale. Note that the flattening off of the Shannon--Jaynes entropy towards vanishing $\rho$.}\label{Fig:cmpreg}
\end{figure}

Let me focus here on one recent Bayesian approach, the~BR method, presented in~\cite{Burnier:2013nla} with regulator \eqref{eq:BR}, which was designed with the particular one-dimensional inverse problem of \eqref{Eq:convdisc} in mind. The~motivation to develop this new method on the one hand arose from the observation that  the specific form of the Shannon--Jaynes regulator of the MEM can pose a problem in finding the optimal Bayesian solution. Consider the (negative of the) integrands of \eqref{eq:TK}--\eqref{eq:BR}, plotted for an arbitrary choice of $\alpha=0.1$ and $m=1$ in the left panel of Figure~\ref{Fig:cmpreg}. By~construction, all of them have an extremum at $\rho=m$ and, as expected, only the Tikhonov regulator allows $\rho$ to take on values smaller than zero. Both the MEM and BR regulator diverge as $\rho\to\infty$, but their behavior close to $\rho=0$ differs markedly. Let us have a closer look in the right panel of Figure~\ref{Fig:cmpreg}. Plotted in log-log scale, we see that while the BR regulator diverges as $\rho\to0$, the Shannon--Jaynes entropy just flattens off and intercepts the \emph{y}-axis at a finite value. It is this flattening off that, in practice, can lead to very slow convergence of the deployed optimization algorithms, as~the MEM wanders about in this flat~direction.

The second reason was that the MEM originally arose in the context of two-dimensional astronomical image reconstruction and the assumptions that enter its construction make reference specifically to this two-dimensional nature of the inverse problems. Here, instead we are interested in a simple one-dimensional inverse problem, which is not directly related to at least one of the axioms underlying the MEM. As~laid out in detail in~\cite{Burnier:2013nla}, we started by replacing that axiom by a generic smoothness axiom and also introduced a scale invariance axiom to arrive at the BR method with its regulator given in \eqref{eq:BR}. The~form of this regulator differs in important aspects from the Shannon--Jaynes entropy. Note that it contains only ratios of the functions $\rho$ and $m$. Since both quantities carry the same units, it implies that the value of the integral does not depend on the units assigned to them. In~contrast, the Shannon--Jaynes integrand, and thus the integral, depend on the specific choice of units in $\rho$. In~addition, the logarithmic term in ${\cal S}_{\rm BR}$ is not multiplied with the function $\rho$. This changes the behavior of the integrand for $\rho\to 0$, making it diverge---i.e., the BR regulator avoids the flat direction encountered in ${\cal S}_{\rm MEM}$ and thus shows much better convergence properties for functions $\rho$, which contain large ranges, where their values are parametrically much smaller than in the dominant~contributions. 

A straightforward implementation of the BR method in a general Bayesian context has recently been introduced in~\cite{Rothkopf:2019dzu}. Using the \href{http://mc-stan.org/}{MC-STAN Monte-Carlo library}, it has been shown how to sample the posterior distribution of the BR method in the full $N_\omega$-dimensional search space and thus how to access the full information encoded in it. The~maximum a-posteriori solution considered in the literature so far only captures the maximum of this distribution. Not only does a full Bayesian implementation of the BR method allow for a self-consistent treatment of the hyperparameter $\alpha$, but it also provides the complete uncertainty budget from the spread of the~posterior.

\section{Summary and~Conclusions}

We have critically assessed in this paper the arguments underlying Bryan's MEM, which we show to be flawed as they resort to methods of linear algebra when treating an inherently nonlinear optimization problem. Therefore, we conclude that even though the individual steps in the derivation of the SVD subspace are all correct, they do not apply to the problem at hand and their conclusions can be disproved with a direct counterexample. The~counterexample we provided utilizes the fact that the componentwise exponentiated columns of the matrix $\hat U$ do not project into the null-space of the Kernel when computing synthetic data. After~establishing the fact that the restriction to the SVD subspace is an ad-hoc procedure, we discussed possible ways to overcome it, suggesting either to systematically extend the search space within the MEM or abandon the MEM in favor of one of the many modern Bayesian approaches developed over the past two~decades.

In our ongoing work to improve methods for the Bayesian treatment of inverse problems, we focus on the development of more specific regulators. So far, the methods described above only make reference to very generic properties of the quantity $\rho$, such as smoothness and positivity. As~a specific example, in~the study of spectral functions in strongly-correlated quantum systems, the~underlying first principles theory provides extra domain knowledge about admissible structures in $\rho$ that so far is not systematically exploited. A~further focus of our work is the extension of the separable regulators discussed in this work to fully correlated prior distributions that exploit cross-correlation between the individual parameters $\rho_l$.

\vspace{6pt} 



\authorcontributions{The author has conceptualized and worked out the argumentative chain, as~well as the explicit examples in the present study. He has written the manuscript and prepared all~figures. The author has read and agreed to the published version of the manuscript.}

\funding{The author acknowledges funding by the Research Council of Norway under the FRIPRO Young Research Talent grant~286883.}


\conflictsofinterest{The author declares that he has no competing~interests.} 

\abbreviations{The following abbreviations are used in this manuscript:\\

\noindent 
\begin{tabular}{@{}ll}
BFGS& Broyden–Fletcher–Goldfarb–Shanno\\
BR & Bayesian Reconstruction\\
MEM& Maximum Entropy Method\\
SJ&Shannon--Jaynes\\
SVD& singular value decomposition\\
TK& Tikhonov
\end{tabular}}

\appendixtitles{yes} 
\appendix

\pagebreak
\clearpage 
\pagebreak

\section{Explicit SVD Example of (Non-) Projection to the Null-Space}
\label{sec:SVDex}

\tikz[remember picture,overlay] \node[opacity=1,inner sep=0pt] at (current page.center){\includegraphics[width=0.9\paperwidth,clip=true, trim=0 2cm 0 0]{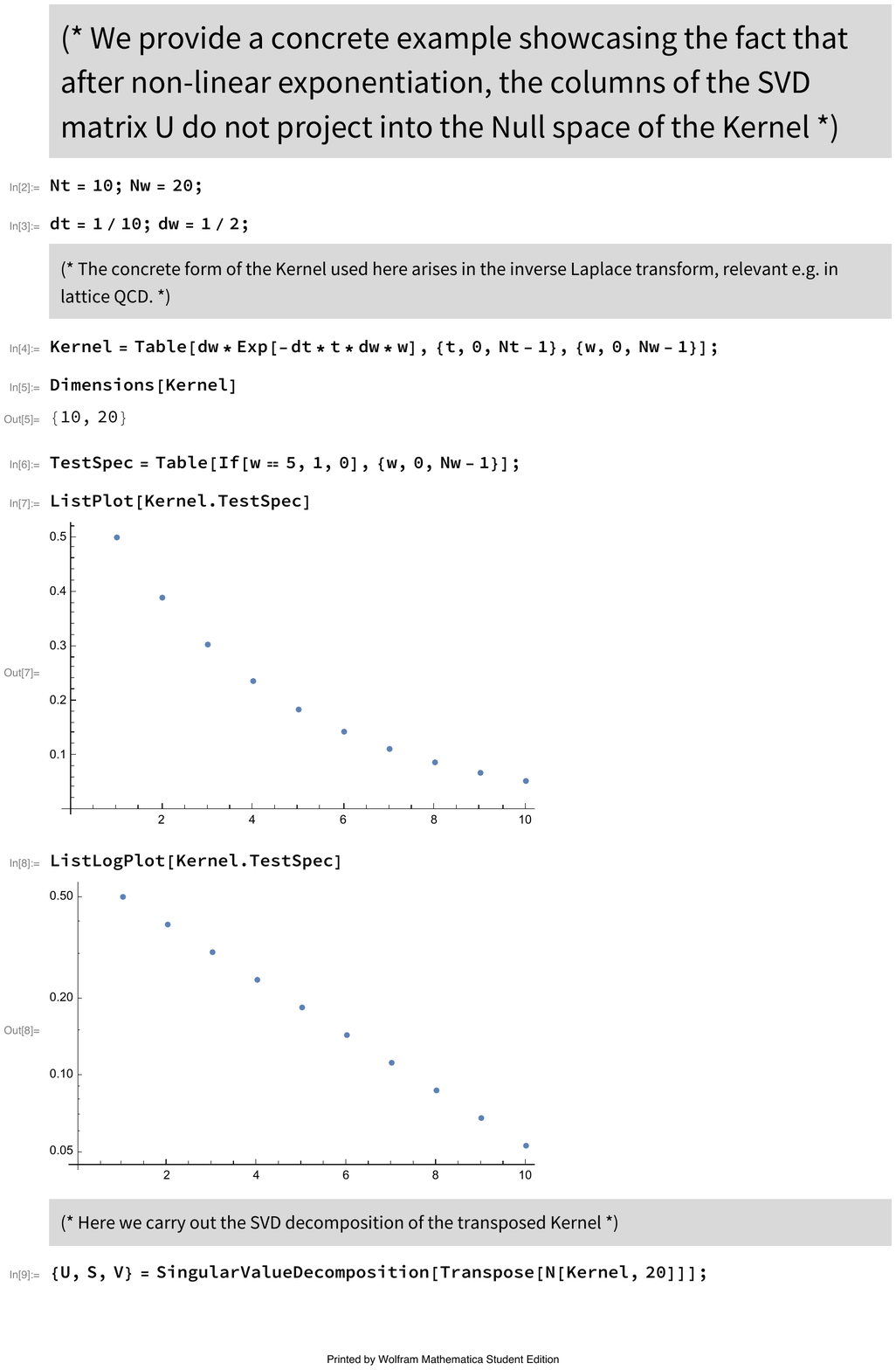}};

\clearpage
\pagebreak

\tikz[remember picture,overlay] \node[opacity=1,inner sep=0pt] at (current page.center){\includegraphics[width=0.9\paperwidth,clip=true, trim=0 2cm 0 2cm]{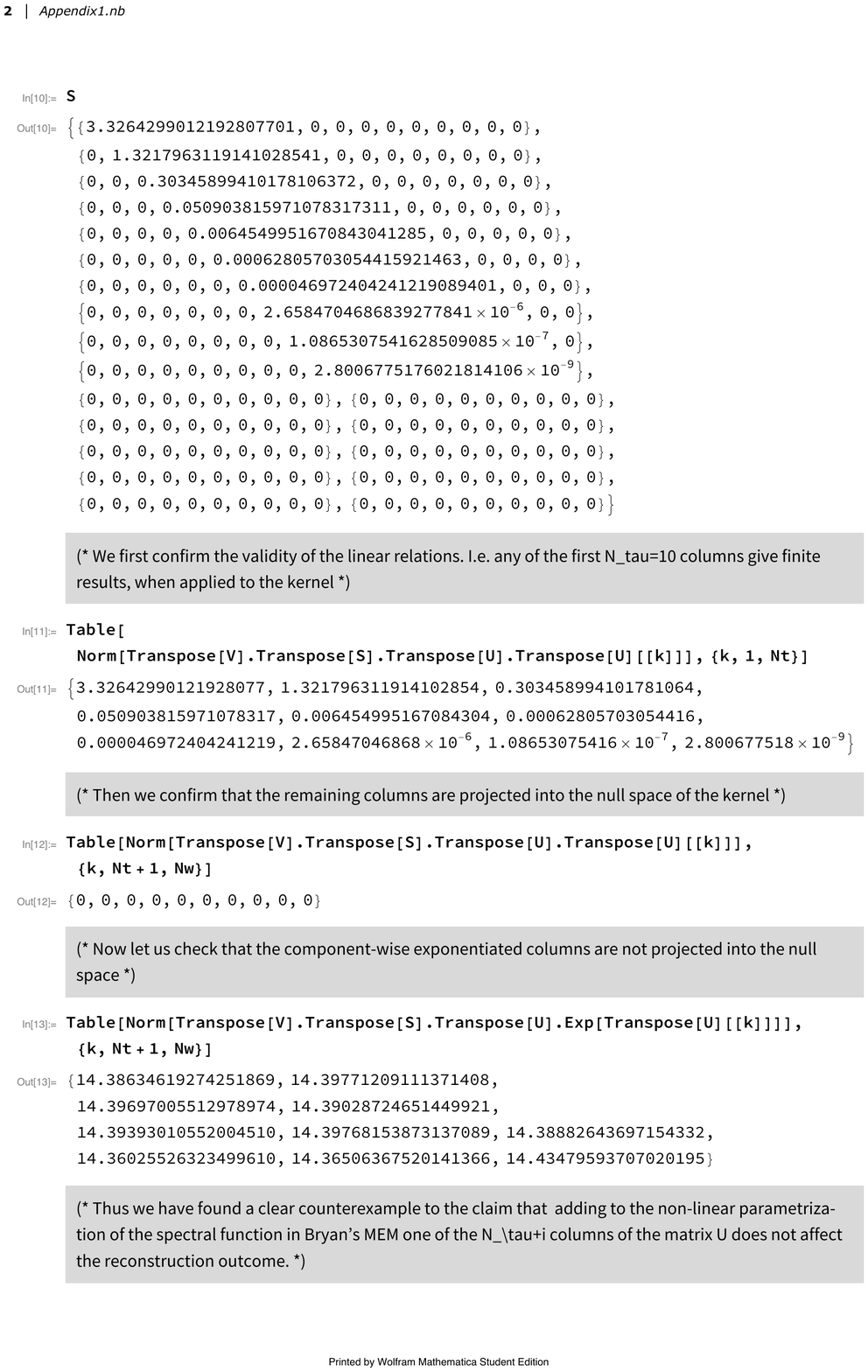}};

\pagebreak
\clearpage 
\pagebreak

\section{Explicit MEM Example with a Solution Outside of the SVD Subspace}
\label{sec:MEMex}

\tikz[remember picture,overlay] \node[opacity=1,inner sep=0pt] at (current page.center){\includegraphics[width=0.9\paperwidth,,clip=true, trim=0 2cm 0 0]{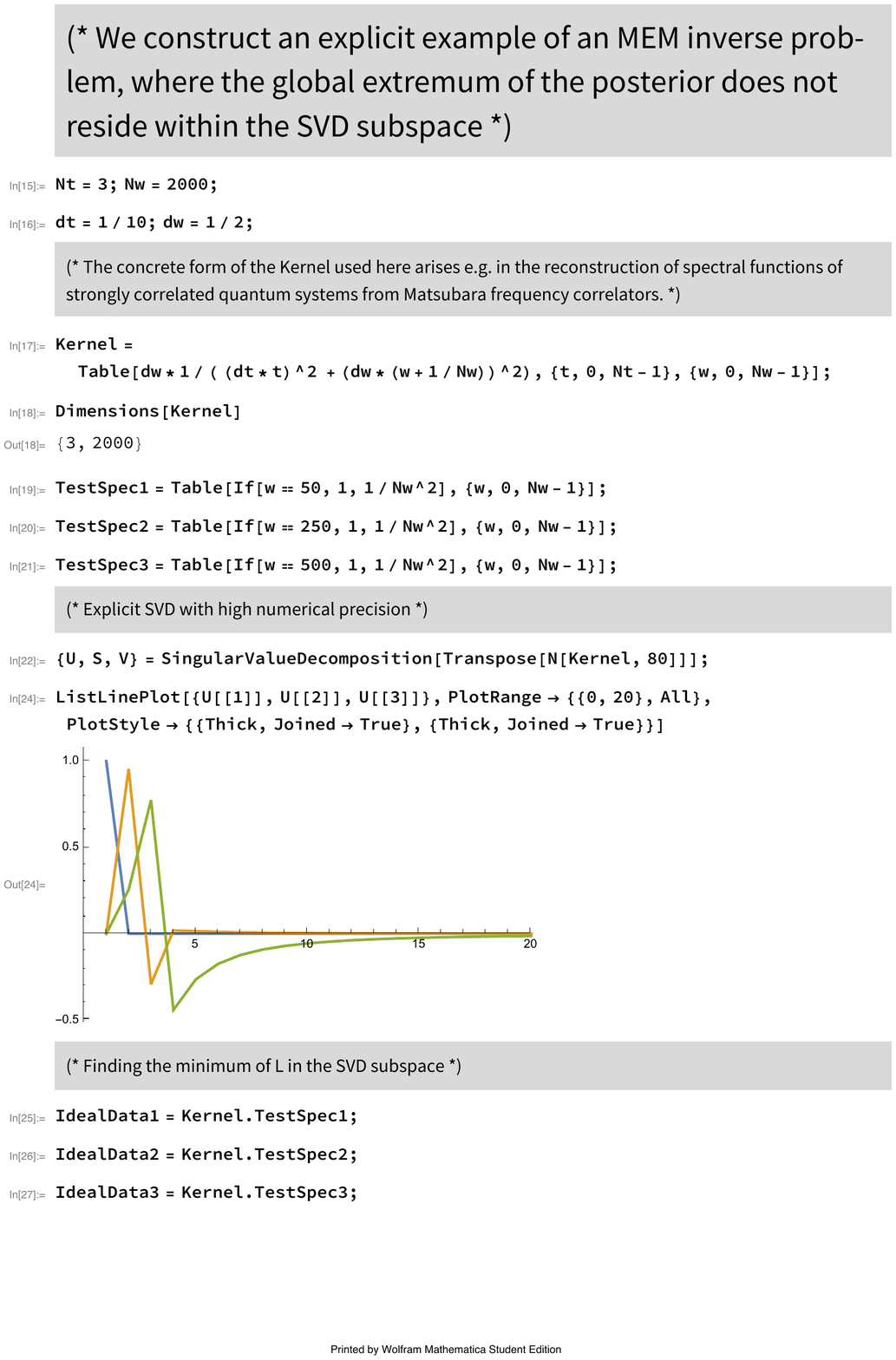}};

\clearpage
\pagebreak

\tikz[remember picture,overlay] \node[opacity=1,inner sep=0pt] at (current page.center){\includegraphics[width=0.9\paperwidth,clip=true, trim=0 2cm 0 2cm]{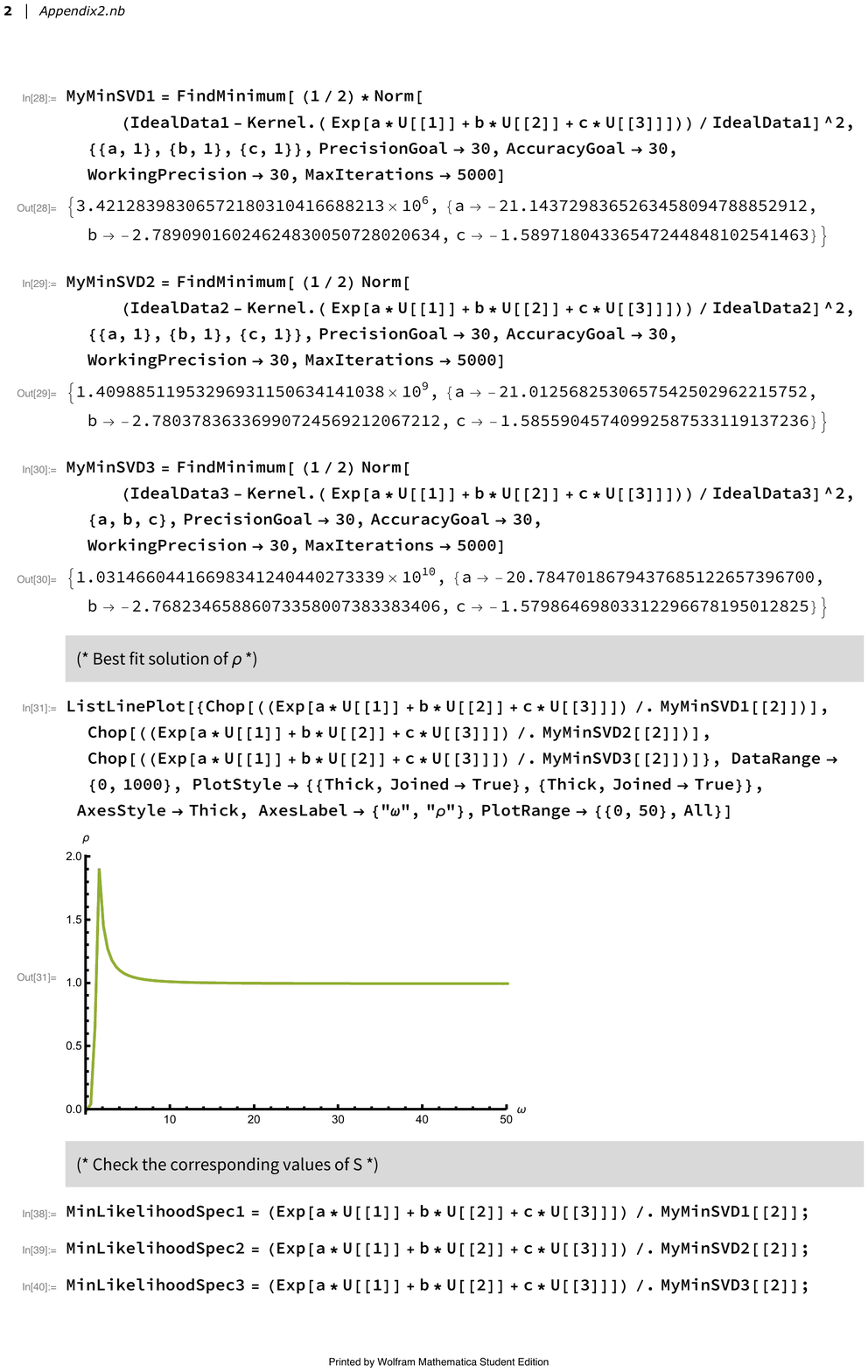}};

\clearpage
\pagebreak

\tikz[remember picture,overlay] \node[opacity=1,inner sep=0pt] at (current page.center){\includegraphics[width=0.9\paperwidth,clip=true, trim=0 2cm 0 2cm]{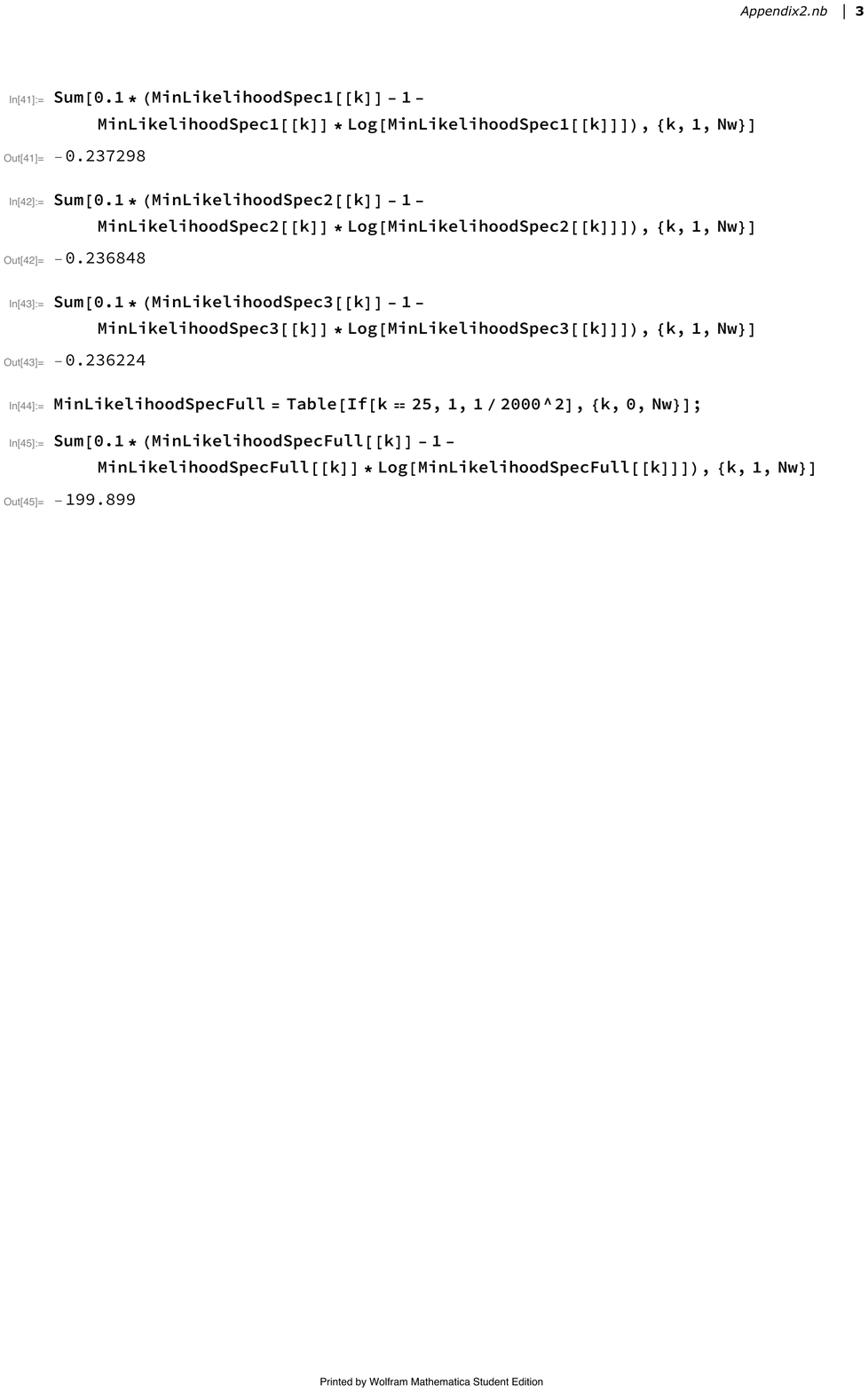}};

\vspace{+240pt}

\reftitle{References}

\end{document}